\documentclass[aip,jcp,reprint,amsmath,amssymb,amsfonts,showkeys,showpacs,preprintnumbers]{revtex4-1}
\usepackage{bm,graphicx,xcolor,hyperref,lineno,multirow,microtype}
\newcommand{\eps}{\varepsilon}

\newcommand{\la}{\langle}
\newcommand{\ra}{\rangle}
\newcommand{\ha}{\frac{1}{2}}
\newcommand{\tha}{\frac{3}{2}}
\newcommand{\mc}{\multicolumn}
\newcommand{\mr}{\multirow}
\newcommand{\Ec}{\varepsilon_{\rm C}}

\begin{document}

\title{Thinking outside the box: the uniform electron gas on a hypersphere}

\author{Pierre-Fran\c{c}ois Loos}
\email{loos@rsc.anu.edu.au}
\author{Peter M. W. Gill}
\thanks{Corresponding author}
\email{peter.gill@anu.edu.au}
\affiliation{Research School of Chemistry, 
Australian National University, Canberra, ACT 0200, Australia}
\date{\today}

\begin{abstract}
We discuss alternative homogeneous electron gas systems in which a finite number $n$ of electrons are confined to a $D$-dimensional sphere.   We derive the first few terms of the high-density ($r_s\to0$, where $r_s$ is the Seitz radius) energy expansions for these systems and show that, in the thermodynamic limit ($n\to\infty$), these terms become identical to those of $D$-dimensional jellium.  
\end{abstract}

\maketitle

\section{Introduction}

The $D$-dimensional uniform electron gas (UEG), or $D$-jellium, is the foundation of most density functionals.  It consists of interacting electrons in an infinite volume and in the presence of a uniformly distributed background positive charge.  Traditionally, in its paramagnetic version, the system is constructed by allowing the number $n$ of paired electrons in a $D$-dimensional box of volume $V$ to approach infinity with $\rho = n/V$ held constant. \cite{Vignale, ParrYang}

Using atomic units, the high-density ($r_s\to0$, where $r_s$ is the Seitz radius) expansion of the reduced energy (\textit{i.e.}~energy per electron) of $D$-jellium is 
\begin{equation}
\label{Ejellium}
        \eps(D) = \eps_{\text{T}}(D) + \eps_{\text{X}}(D) + \Ec(D),
\end{equation}
where $\eps_{\text{T}}$ and $\eps_{\text{X}}$ are kinetic \cite{Fermi26, Thomas27} and exchange \cite{Dirac30, Friesecke97} energies
\begin{align}
\label{EHFjellium}
        \eps_{\text{T}}(D)	& = \frac{a_{-2}(D)}{r_s^2},	&	\eps_{\text{X}}(D)	& = \frac{a_{-1}(D)}{r_s},
\end{align}
and $\Ec$ is the correlation energy.  After many important contributions, \cite{Macke50, GellMann57, Onsager66, Stern73, Rajagopal77, Isihara80, Hoffman92, Seidl04, Sun10, 2jellium, 3jellium, Ceperley80, Tanatar89, Kwon93, Ortiz94, Rapisarda96, Kwon98, Ortiz99, Attaccalite02, Drummond09b} it is known that, for 2- and 3-jellium, the correlation energy takes the form
\begin{equation}
\label{Ecjellium}
	\Ec(D) = \sum_{j=0}^\infty \left[a_j(D) + b_j(D)\ln r_s\right]r_s^j.
\end{equation}
The constant term in \eqref{Ecjellium} is usually decomposed as
\begin{equation}
	a_0(D) = a_{0,\text{J}}(D) + a_{0,\text{K}}(D),
\end{equation}
where $a_{0,\text{J}}$ is the direct (``ring-diagram'') contribution, and $a_{0,\text{K}}$ is the second-order exchange part.
The first few $a_j$ and $b_j$ and are known analytically or numerically for the important $D=2$ and $D=3$ cases (see Table \ref{tab:coeffs}).

In this Article, we introduce an alternative paradigm, in which the electrons are confined to a $D$-sphere, that is, the surface of a ($D+1$)-dimensional ball. These systems possess uniform densities, even for finite $n$, and because all points on a $D$-sphere are equivalent, their mathematical analysis is relatively straightforward. \cite{TEOAS09, Quasi09, LoosConcentric, LoosHook, LoosExcitSph}

Electronic properties of the UEG on a 2-sphere have been previously studied in modeling multielectron bubbles in liquid helium (see Ref.~\onlinecite{Tempere07}), and similarities between this system and 2-jellium have been noticed by Longe and Bose. \cite{Longe96}  However, the UEG on a 3-sphere has not been considered before, and this Article presents the first study of correlation effects in a spherically-confined three-dimensional UEG.

\section{Hartree-Fock Energies}

\subsection{Exact results for finite $n$}

The orbitals for an electron on a $D$-sphere of radius $R$ are the normalized hyperspherical harmonics $Y_{\ell\mu}$, where $\ell$ is the principal quantum number and $\mu$ is a composite index of the remaining quantum numbers. \cite{AveryBook, Avery93}  We confine our attention to systems in which every orbital with $\ell = 0,1,\ldots,L$ is occupied by two electrons, thus yielding an electron density that is uniform on the sphere (see Eq.~\eqref{rho} below).  
The resulting model is defined completely by the three parameters $D$, $L$ and $R$.

The volume of a $D$-sphere is 
\begin{equation}
	V = \frac{2\pi^{\frac{D+1}{2}}}{\Gamma\left(\frac{D+1}{2}\right)}R^D,
\end{equation}
where $\Gamma$ is the gamma function, \cite{NISTbook} the number of orbitals with quantum number $\ell$ is
\begin{equation}
\label{nl}
	n_\ell = \frac{(2\ell+D-1) \Gamma(\ell+D-1)}{\Gamma(D) \Gamma(\ell+1)},
\end{equation}
and each of these has energy 
\begin{equation}
	\kappa_\ell = \frac{\ell(\ell+D-1)}{2R^2}.
\end{equation}
Because the total number of electrons is
\begin{equation} 
\label{n}
	n = 2\sum_{\ell=0}^{L} n_\ell 
	= 2\frac{(2L+D) \Gamma(L+D)}{\Gamma(D+1) \Gamma(L+1)},
\end{equation}
it follows that the uniform electron density is
\begin{equation}
\label{rho}
\begin{split}
	\rho 
	& = \frac{n}{V} 
	= \frac{\Gamma(D/2+1)}{\pi^{D/2} r_s^D} 
	\\
	& = \frac{(2L+D)\Gamma(L+D)}{\Gamma(D/2+1)\Gamma(L+1)} \frac{1}{(4\pi R^2)^{D/2}},
\end{split}
\end{equation}
and the Seitz radius is
\begin{equation}
	r_s = \alpha_D \left[
	\frac{\Gamma(L+1)}
	{(L+D/2)\Gamma(L+D)}\right]^{1/D} R,
\end{equation}
with
\begin{equation}
	\alpha_D = 2^{1-1/D}\Gamma^{2/D}(D/2+1).
\end{equation}

\begin{figure}
	\begin{center}
	\includegraphics[width=0.4\textwidth]{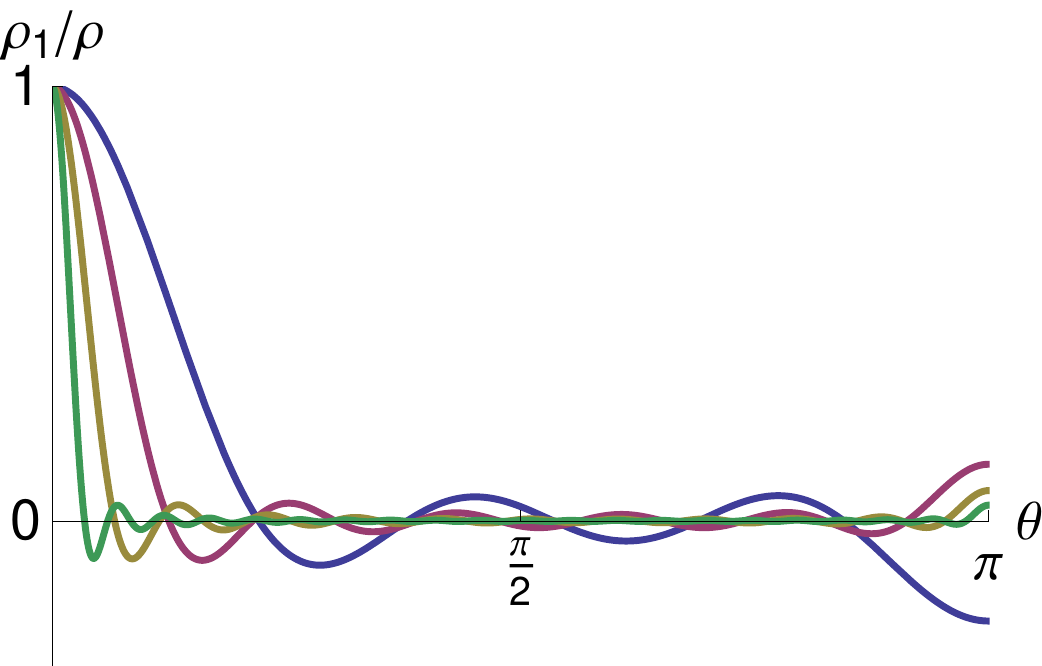}
	\caption{
	\label{fig:myopia}
	The one-particle density matrix for the UEG on a 3-sphere.
	Plots for $L=5$ (blue), $L=10$ (red), $L=20$ (yellow) and $L=40$ (green).}
	\end{center}
\end{figure}

Using the hyperspherical harmonic addition theorem, \cite{Avery93} one finds that the one-particle density matrix is
\begin{equation}
\begin{split}
\label{rho1}
	\rho_1(\bm{\Omega}_1,\bm{\Omega}_2)
	& = 2 \sum_{\ell=0}^{L} 
	\sum_{\mu=1}^{n_\ell}
	Y_{\ell\mu}^*(\bm{\Omega}_1)Y_{\ell\mu}(\bm{\Omega}_2)
	\\
	& = \rho \frac{\Gamma\left(D/2+1\right) \Gamma\left(L+1\right)}{\Gamma\left(L+D/2+1\right)}P_{L}^{(D/2,D/2-1)}(\cos \theta) ,
\end{split}
\end{equation}
where $P_L^{(\alpha,\beta)}$ is a $L$th degree Jacobi polynomial. \cite{NISTbook} The angle $\theta$ is that subtended by the electrons at the origin and is related to the interelectronic distance by the relation \footnote{In our model, the electrons interact ``through'' the sphere.  Thus, $r_{12}$ is the length of the line, not the arc, connecting the electrons.}
\begin{equation}
	r_{12} \equiv \left| \bm{r}_1 - \bm{r}_2 \right| = 2R\sin(\theta/2).
\end{equation}
The density matrix  decays rapidly with interelectronic separation when $L$ is large (Fig. \ref{fig:myopia}), illustrating the ``short-sightedness'' of matter. \cite{Kohn96, Kohn05}

Many properties of the UEG on a $D$-sphere can be found from Eqs. \eqref{nl} -- \eqref{rho1}.  Its kinetic energy, for example, is
\begin{equation} \label{ET}
	\eps_\text{T}(D,L) = \frac{2}{n} \sum_{\ell=0}^L n_\ell \kappa_\ell = \frac{D}{2(D+2)}\frac{L(L+D)}{R^2},
\end{equation}
and it can be shown that its exchange energy is
\begin{equation} \label{EX}
\begin{split}
	\eps_\text{X}(D,L)
		= & -\frac{1}{2n} \iint \frac{\rho_1(\bm{\Omega}_1,\bm{\Omega}_2)^2}{r_{12}} \ d\bm{\Omega}_1\,d\bm{\Omega}_2
		\\
		= & -\frac{n}{2R} \frac{\Gamma(\frac{D+1}{2})}{\sqrt\pi\,\Gamma(\frac{D}{2})}
			\frac{D\,\Gamma(D-1)}{2L+D} \frac{\Gamma(L+\tha)}{\Gamma(L+D-\ha)}
		\\
		& \times {_4F_3} \left[\begin{array}{cc}
		\begin{array}{cccc}
			-L,	&	L+D,	&	\frac{D-1}{2},	&	-\ha	
		\end{array}	
		\\
		\begin{array}{ccc}
			-L-\ha,	&	L+D-\ha,	&	\frac{D+2}{2}	
		\end{array}
		\end{array}	;1 \right],
\end{split}
\end{equation}
where $_4F_3$ is a generalized hypergeometric function. \cite{NISTbook}

\begin{table}
\caption{\label{tab:coeffs} 
High-density coefficients for $D$-jellium and the UEG on a $D$-sphere. $\beta$ and $\zeta$ are the Dirichlet beta and Riemann zeta functions. \cite{NISTbook}}
\begin{ruledtabular}
\begin{tabular}{ccccc}
	Coefficient			&	Term					&						&			$D=2$					&				$D=3$				\\
	\hline
	$a_{-2}$				&	$r_s^{-2}$				&						&	$1/2$							&	$\frac{3}{10}(9\pi/4)^{2/3}$			\\
	$a_{-1}$				&	$r_s^{-1}$				&						&	$-\frac{4\sqrt{2}}{3\pi}$				&	$-\frac{3}{4\pi}(9\pi/4)^{1/3}$			\\
	$b_0$				&	$\ln r_s$				&						&	$0$								&	$(1-\ln2)/\pi^2$						\\
	\mr{2}{*}{$a_0$}		&	\mr{2}{*}{$r_s^0$}		&	$a_{0,\text{J}}$			&	$\ln 2 -1$							&	$-0.071099$						\\
						&						&	$a_{0,\text{K}}$			&	$\beta(2)-\frac{8}{\pi^2}\beta(4)$		&	 $\frac{\ln 2}{6}-\frac{3}{4\pi^2} \zeta(3)$	\\
	$b_1$				&	$r_s\ln r_s$			&						&	$-\sqrt{2}\left(\frac{10}{3\pi}-1\right)$		&	$+0.009229$						\\
\end{tabular}
\end{ruledtabular}
\end{table}

\subsection{The thermodynamic limit}

The above expressions are exact for all $L$ but, in the thermodynamic limit ($n,L\to\infty$), each simplifies significantly.
For example,
\begin{align}
	n	& \to \frac{4}{\Gamma(D+1)} L^D,
	\label{nlim}
	\\
	\rho	& \to \frac{2}{\Gamma(D/2+1)}\frac{L^D}{(4\pi R^2)^{D/2}},
	\label{rholim}	
	\\
	r_s	&	\to	\alpha_D\frac{R}{L},
	\label{rslim}	
	\\
	\rho_1	& \to \rho\ 
	\frac{\Gamma(D/2+1) J_{D/2}(L\theta)}{(L\theta/2)^{D/2}},
	\label{rho1lim}
\end{align}
where $J_n$ is the $n$th-order Bessel function. \cite{NISTbook}  We note that \eqref{rho1lim} reduces to the usual density matrices in 2-jellium \cite{Glasser77} and 3-jellium. \cite{Dirac30}  The kinetic and exchange energies become
\begin{align}
	\eps_\text{T}(D)	& = +\frac{D}{2(D+2)}\frac{\alpha_D^2}{r_s^2},
	\label{ETlim}
	\\
	\eps_\text{X}(D)	& = -\frac{2D}{\pi(D^2-1)}\frac{\alpha_D}{r_s}.
	\label{EXlim}
\end{align}
Equations \eqref{ETlim} and \eqref{EXlim} yield the two terms in \eqref{EHFjellium}, and are identical to the $D$-jellium expressions.
Particular cases are given in Table \ref{tab:coeffs}.  These results were originally discovered by Glasser and Boersma, \cite{Glasser83} and Iwamoto \cite{Iwamoto84} for $D$-jellium, but our derivation for the UEG on a $D$-sphere is more compact than theirs.

\begin{table}
\caption{
\label{tab:finiteL}
Numerical values of $a_{0,\text{J}}(2,L)$, $a_{0,\text{K}}(2,L)$, $c_{0,\text{J}}(3,L)$ and $a_{0,\text{K}}(3,L)$ for various $L$.}
\begin{ruledtabular}
\begin{tabular}{cccccccc}
	$L$	&						\mc{3}{c}{UEG on a 2-sphere}					&&					\mc{3}{c}{UEG on a 3-sphere}						\\
									\cline{2-4}													\cline{6-8}
		& $n$	&	$a_{0,\text{J}}(2,L)$	&	$a_{0,\text{K}}(2,L)$	&& $n$	&	$c_{0,\text{J}}(3,L)$	&	$a_{0,\text{K}}(3,L)$	\\
	\hline
	0	&	2	&		$-0.2274$			&	$+0.1137$				&&	2	&		$-0.0476$			&	$+0.0238$				\\
	1	&	8	&		$-0.2534$			&	$+0.1111$				&&	10	&		$-0.0717$			&	$+0.0231$				\\
	2	&	18	&		$-0.2677$			&	$+0.1118$				&&	28	&		$-0.0897$			&	$+0.0231$				\\
	3	&	32	&		$-0.2762$			&	$+0.1124$				&&	60	&		$-0.1038$ 		&	$+0.0233$				\\
	4	&	50	&		$-0.2816$			&	$+0.1128$				&&	110	&		$-0.1154$			&	$+0.0234$				\\
	\vdots	&	\vdots	&	\vdots		&	\vdots					&&	\vdots		&\vdots			&	\vdots					\\
	$\infty$	&	$\infty$	&	$-0.3069$		&	$+0.1144$				&&	$\infty$		&$-\infty$			&	$+0.0242$				\\
\end{tabular}
\end{ruledtabular}
\end{table}

\begin{figure}
	\begin{center}
	\includegraphics[width=0.4\textwidth]{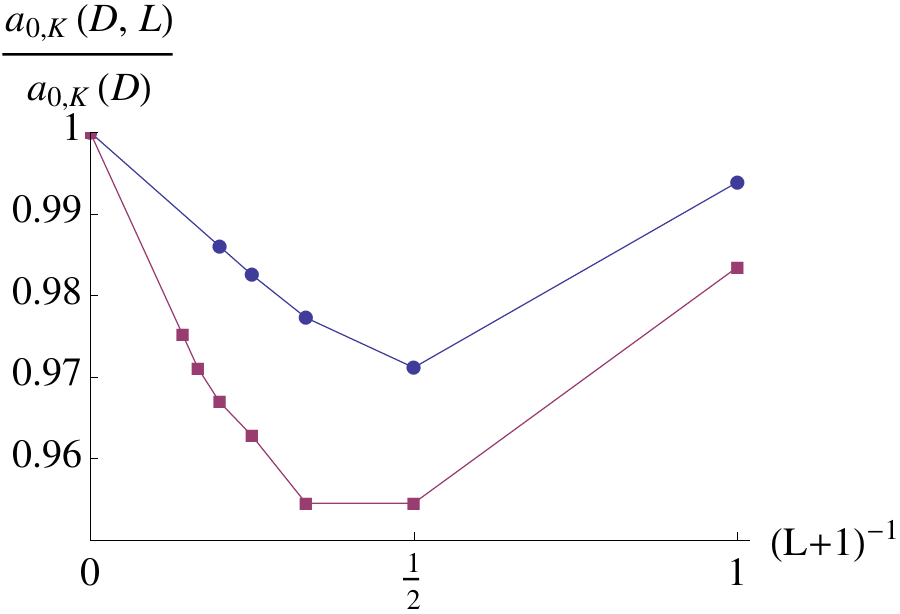}
	\caption{
	\label{fig:finiteL}
	$\frac{a_{0,\text{K}}(D,L)}{a_{0,\text{K}}(D)}$ as a function of $(L+1)^{-1}$ for $D=2$ (blue dots) and $D=3$ (red squares) .}
	\end{center}
\end{figure}

\section{Correlation energies}

\subsection{Exact results for finite $n$}

We now turn our attention to the study of the correlation energy of the spherically-confined UEGs.  By applying perturbation theory to UEG on a 2-sphere, we find that the reduced energy coefficient corresponding to the lowest-order ring-diagram contribution is
\begin{equation}
\label{eps0D-2D-3j}
\begin{split}
	a_{0,\text{J}}(2,L)
		= & \frac{2}{n} \sum_{i,j}^\text{occ} \sum_{a,b}^\text{virt} \frac{\la ij |ab \ra^2}{\kappa_i+\kappa_j-\kappa_a-\kappa_b}
		\\
		= & \frac{1}{n}\sum_{i,j=0}^{L} \sum_{a,b=L+1}^\infty \frac{(2i+1)(2j+1)(2a+1)(2b+1)}{\kappa_i+\kappa_j-\kappa_a-\kappa_b}
		\\
		& \times
		\sum_{\ell} \frac{1}{2\ell+1}
			\begin{pmatrix}
				i	&	\ell	&	a	\\
				0	&	0	&	0	\\
			\end{pmatrix}^2
			\begin{pmatrix}
				j	&	\ell	&	b	\\
				0	&	0	&	0	\\
			\end{pmatrix}^2,
\end{split}
\end{equation}
where $\la ij |ab \ra$ are two-electron integrals and the brackets are 3j symbols. \cite{NISTbook}   For the UEG on a 3-sphere, the coupling coefficient in SO(4) is much simpler than in SO(3) \cite{Alisauskas02} and the energy coefficient from the lowest-order ring-diagram is
\begin{multline}
\label{eps0D-3D-3j}
	c_{0,\text{J}}(3,L) 
	= \frac{1}{n}\sum_{i,j=0}^{L} \sum_{a,b=L+1}^{\infty} 
	\frac{(i+1)(j+1)(a+1)(b+1)}
	{\kappa_i+\kappa_j-\kappa_a-\kappa_b}  
	\\
	\times
	\sum_{\ell} \frac{(2/\pi)^2}{(\ell+\ha)^2(\ell+\tha)^2},
\end{multline}
where the sum over $\ell$ respects the same restrictions as in the 3j symbols in \eqref{eps0D-2D-3j}.  

The second-order exchange part for the UEG on a 2-sphere is
\begin{equation}
\begin{split}
\label{b0-x-2D}
	a_{0,\text{K}}(2,L)
		= & \frac{1}{n} \sum_{i,j}^\text{occ} \sum_{a,b}^\text{virt} \frac{\la ij |ab \ra\la ba |ij \ra}{\kappa_a+\kappa_b-\kappa_i-\kappa_j}
		\\
		= & \frac{1}{n} \sum_{i,j=0}^L \sum_{a,b=L+1}^{\infty} \frac{(2i+1)(2j+1)(2a+1)(2b+1)}{\kappa_a+\kappa_b-\kappa_i-\kappa_j}
		\\
		& \times \sum_{\ell_1,\ell_2} (-1)^{\ell_1+\ell_2}
			\begin{pmatrix}
				i	&	\ell_1	&	a	\\
				0	&	0	&	0		\\
			\end{pmatrix}
			\begin{pmatrix}
				j	&	\ell_1	&	b	\\
				0	&	0	&	0		\\
			\end{pmatrix}
			\\
			& \times
			\begin{pmatrix}
				i	&	\ell_2	&	b	\\
				0	&	0	&	0	\\
			\end{pmatrix}
			\begin{pmatrix}
				j	&	\ell_2	&	a	\\
				0	&	0	&	0	\\
			\end{pmatrix}
			\begin{Bmatrix}
				i	&	\ell_1	&	a	\\
				j	&	\ell_2	&	b	\\
			\end{Bmatrix},
\end{split}
\end{equation}
where the curly brackets denote 6j symbols, \cite{NISTbook} and for the UEG on a 3-sphere, we found
\begin{multline}
\label{b0-x-3D}
	a_{0,\text{K}}(3,L)
	= \frac{1}{n} \sum_{i,j=0}^{L} \sum_{a,b=L+1}^{\infty} 
	\frac{(i+1)(j+1)(a+1)(b+1)}
	{\kappa_a+\kappa_b-\kappa_i-\kappa_j} 
	\\
	\times
	\sum_{\ell_1,\ell_2} 
	\frac{(2/\pi)(\ell_1+1)}{(\ell_1+\ha)(\ell_1+\tha)}
	\frac{(2/\pi)(\ell_2+1)}{(\ell_2+\ha)(\ell_2+\tha)}
	\begin{Bmatrix}
		i	&	\ell_1	&	a	\\
		j	&	\ell_2	&	b	\\
	\end{Bmatrix},
\end{multline}
where we have used the SO(4) version of the 6j symbols. \cite{Alisauskas02}  Numerical values for finite $L$ are given in Table \ref{tab:finiteL}.

\subsection{The thermodynamic limit}

In the thermodynamic limit ($n\to\infty$), Eq. \eqref{eps0D-2D-3j} becomes
\begin{multline}
	a_{0,\text{J}}(2)
	= - \int_{0}^{\infty} \frac{d\ell}{\ell} 
	\int_0^1 i\,di \int_0^1 j\,dj 
	\int_{\max\left(1,\left|\ell-i\right|\right)}^{\ell+i} a\,da
	\\ \times
	\int_{\max\left(1,\left|\ell-j\right|\right)}^{\ell+j} b\,db
	\frac{\mathcal{J}_{i,\ell,a} \mathcal{J}_{j,\ell,b}}{a^2+b^2-i^2-j^2},
\end{multline}
where
\begin{equation}
	\mathcal{J}_{i,\ell,a} = \frac{2}{\pi}\frac{1}{\sqrt{(i+\ell+a)(i+\ell-a)(\ell+a-i)(a+i-\ell)}}
\end{equation}
comes from the asymptotic expansion of the 3j symbol. \cite{Borodin78}  Defining $a^2=i^2+\ell^2-2i\ell \cos \theta_1$ and $b^2=j^2+\ell^2-2j\ell \cos \theta_2$ to transform the $a$ and $b$ integrals into $\theta_1$ and $\theta_2$ integrals, and then renaming $i$, $j$ and $\ell$ as $p_1$, $p_2$ and $q$, we obtain
\begin{multline}
\label{E2d-2D-Final}
	a_{0,\text{J}}(2)
	= - \frac{1}{4\pi^3}
	\int \frac{d\bm{q}}{q^2}
	\\ \times
	\int_{\substack{\left|\bm{p}_1\right|<1\\\left|\bm{p}_1+\bm{q}\right|>1}}
	\int_{\substack{\left|\bm{p}_2\right|<1\\\left|\bm{p}_2+\bm{q}\right|>1}} 
	\frac{d\bm{p}_1 d\bm{p}_2}{q^2+\bm{q}\cdot\left(\bm{p}_1+\bm{p}_2\right)},
\end{multline}
which is precisely the lowest-order ring-diagram contribution of 2-jellium. \cite{Rajagopal77}  This integral can be solved \cite{2jellium} to yield
\begin{equation}
	a_{0,\text{J}}(2) = \ln 2 - 1.
\end{equation}
One also finds that the higher-order ring-diagram contributions are identical in 2-jellium and in the UEG on a 2-sphere and the resummation technique \cite{Rajagopal77} yields 
\begin{align}
	b_{0}(2) & = 0,  & b_{1}(2) & = -\sqrt{2}\left(\frac{10}{3\pi}-1\right).
\end{align}

For large $L$, the sums in Eq. \eqref{eps0D-3D-3j} can be replaced by integrals and one finds and the leading order of \eqref{eps0D-3D-3j} is
\begin{equation}
\label{Div}
\begin{split}
	c_{0,\text{J}}(3,L) 
	& \sim - \frac{3}{\pi^2} \int_{1/L}^{1} d\ell \int_0^{\infty} dt
	\\
	& \quad  \times
	\left[ \int_{1-\ell}^{1} \int_{1}^{i+\ell} \frac{ai}{\ell^2} e^{-(a^2-i^2)t} da\,di\right]^2 
	\\
	& = \frac{1-\ln 2}{\pi^2} \ln \frac{1}{L} + O\left(L^0\right) 
	\\
	& = \frac{1-\ln 2}{\pi^2} \ln r_s + O\left(r_s^0\right).
\end{split}
\end{equation}
It follows that 
\begin{equation}
	b_{0}(3) = \frac{1-\ln 2}{\pi^2},
\end{equation}
and thus the logarithmic divergence of the correlation energy in the UEG on a 3-sphere is exactly the same as in 3-jellium.  One notes that the result \eqref{Div} can be derived for any value of the radius $R$.  The latter divergence, contrary to some claims in the literature, does not result from the long-range part of the Coulomb operator but from its short-range part. \cite{Zecca04, Paziani06}  The observation of the same divergence in the UEG on a 3-sphere --- where the interelectronic distance can never exceed $2R$ --- also demonstrates this.

Proceeding similarly to the $D=2$ case, it can be shown that, in the thermodynamic limit, \eqref{eps0D-3D-3j} becomes identical to the expression of the second-order ring-diagram in 3-jellium:
\begin{multline}
\label{E2d-3D-Final}
	c_{0,\text{J}}(3)
	= - \frac{3}{16\pi^5}
	\int \frac{d\bm{q}}{q^4}
	\\
	\times \int_{\substack{\left|\bm{p}_1\right|<1\\\left|\bm{p}_1+\bm{q}\right|>1}}
	\int_{\substack{\left|\bm{p}_2\right|<1\\\left|\bm{p}_2+\bm{q}\right|>1}}
	\frac{d\bm{p}_1 d\bm{p}_2}{q^2+\bm{q}\cdot\left(\bm{p}_1+\bm{p}_2\right)},
\end{multline}
where the excitation vector $q$ has the domain $ \sqrt{r_s} < q < \infty$. \cite{GellMann57}  Moreover, the higher-order ring diagram contributions are also identical in 3-jellium and in the UEG on a 3-sphere.  Using the resummation technique,  \cite{GellMann57, Hoffman92}  it follows that \footnote{There exists some disagreement in the literature over the value of $\eps_{0,\text{J}}(3)$.  See Ref.~\onlinecite{Porter09} for more details.}
\begin{equation}
	a_{0,\text{J}}(3) = -0.071099.
\end{equation} 

For $D=2$ and $3$, we have not been able to prove the equivalence of the second-order exchange contributions in $D$-jellium and in the UEG on a $D$-sphere, but the numerical results in Table \ref{tab:finiteL}  and Fig.~\ref{fig:finiteL} suggest that, in the thermodynamic limit, $a_{0,\text{K}}(2) \approx +0.11$ and $a_{0,\text{K}}(3) \approx +0.024$, which may be compared with the known 2-jellium and 3-jellium values: \cite{Isihara80, Onsager66}
\begin{align}
	a_{0,\text{K}}(2) & = \beta(2)-\frac{8}{\pi^2}\beta(4) = +0.114357,
	\\
	 a_{0,\text{K}}(3) & = \frac{\ln 2}{6} - \frac{3}{4\pi^2}\zeta(3) = +0.024179,
\end{align}
where $\beta$ and $\zeta$ are the Dirichlet beta and Riemann zeta functions. \cite{NISTbook}

\section{Discussion}

Uniform electron gases on a $D$-sphere are an attractive generalization of $D$-jellium and, as we have shown, one can derive compact expressions for the first few terms of the high-density energy expansions for both finite and infinite systems.  Although UEGs on a $D$-sphere are physically different from $D$-jellium, we have shown that, in the thermodynamic limit, the first few terms of their high-density energy expansions are identical and we conjecture that the high-density expansions are identical to all orders.  

Recent calculations on the Thomson problem suggest that the leading term of the low-density (large-$r_s$) energy expansions in 2-jellium and in the UEG on a 2-sphere are also identical. \cite{Bowick02} Moreover, because the Thomson problem is trivial for $D=1$, it is actually possible to show the strict equality of the leading term of the low-density energy expansions in 1-jellium and in the UEG on a ring (1-sphere). \cite{Fogler05}

Although it is pleasing to know that the spherical and conventional gases become equivalent in the thermodynamic limit, we believe that it is even more important to recognize that they are \emph{not} equivalent for finite $n$.  Equations \eqref{ET} and \eqref{EX} predict significantly different kinetic and exchange energies from \eqref{ETlim} and \eqref{EXlim} when $n$ is small.  Moreover, combining the information from the high- and low-density regimes, one can easily construct local-density approximation-type correlation functionals for finite systems using interpolation functions. \cite{Seidl00, Seidl04, Sun10}  We believe that the UEG on a $D$-sphere will be useful in the future development of correlation functionals within density-functional theory. \cite{UEGs} 

\begin{acknowledgments}
P.M.W.G. thanks the NCI National Facility for a generous grant of supercomputer time and the Australian Research Council (Grants DP0984806 and DP1094170) for funding.
\end{acknowledgments}

\end{document}